\documentclass[11pt]{article}

\usepackage[a4paper,margin=2.35cm]{geometry}
\usepackage{amsmath,amssymb,mathtools}
\usepackage{graphicx}
\usepackage{flafter}
\usepackage{placeins}
\usepackage{subcaption}
\usepackage{booktabs}
\usepackage{array}
\usepackage{microtype}
\usepackage{xcolor}
\usepackage{cite}
\numberwithin{equation}{section}
\usepackage{hyperref}

\hypersetup{
	colorlinks=true,
	linkcolor=black,
	citecolor=black,
	urlcolor=black,
	pdftitle={Complexity Equals Anything for Generalized Born--Infeld AdS Black Holes},
	pdfauthor={H. Babaei-Aghbolagh, Komeil Babaei Velni, M. Reza Mohammadi Mozaffar, Hamid R. Imani}
}

\allowdisplaybreaks
\newcommand{\GN}{G_{\!N}}
\newcommand{\LAdS}{L}
\newcommand{\LAC}{L_{\rm A}}
\newcommand{\lBI}{\lambda_{\mathrm{BI}}}
\newcommand{\aX}{a_X}
\newcommand{\qg}{q}
\newcommand{\rh}{r_h}
\newcommand{\rmin}{r_{\min}}
\newcommand{\PV}{P_v}
\newcommand{\Veff}{U_X}
\newcommand{\dd}{\mathrm{d}}
\newcommand{\ee}{\mathrm{e}}

\title{\bfseries Complexity Equals Anything for Generalized Born--Infeld AdS Black Holes}

\author{
	Hamid R. Imani$^{*}$, H. Babaei-Aghbolagh$^{\dag}$, Komeil Babaei Velni$^{*}$, and \\[-0.05em]
	M. Reza Mohammadi Mozaffar$^{*}$ \\[0.55em]
	\small $^{\dag}$Institute of Fundamental Physics and Quantum Technology, \&\\[-0.1em]
	\small School of Physical Science and Technology, Ningbo University,\\[-0.1em]
	\small Ningbo, Zhejiang 315211, China\\[0.25em]
	\small $^{*}$Department of Physics, University of Guilan, P.O. Box 41335-1914, Rasht, Iran\\[0.65em]
	\small \texttt{hamidrezaimani@webmail.guilan.ac.ir},
	\small \texttt{hosseinbabaei@nbu.edu.cn}, \small\texttt{babaeivelni@guilan.ac.ir},\\[-0.1em]
	\small \texttt{mmohammadi@guilan.ac.ir} 
}

\date{}

\begin{document}
	
	\maketitle
	
	\begin{abstract}
		We study how complexity-equals-anything observables probe the interiors of four-dimensional generalized Born--Infeld AdS black holes with spherical, planar, and hyperbolic horizons. We compare three scalar weights, the on-shell electromagnetic Lagrangian, the negative trace of the stress tensor, and the square of the Weyl tensor. These weights respond differently to the black-hole interior. These weights respond differently in the black hole interior. Specifically, as the type of singularity inside the black hole changes, the three weighted effective potentials change sign from $0^{+}$ to $0^{-}$, from a finite positive value to a finite negative value, and from positive infinity to negative infinity, respectively. This indicates that complexity-equals-anything can probe the type of singularity. We determine the conditions under which extremal hypersurfaces can reach the AdS boundaries and study how their momenta control complexity growth. Our numerical analysis shows that the early-time momentum is predominantly more sensitive to fractional variations in the dressed charge than to equal fractional variations in the Born--Infeld deformation. We also derive an exact relation governing the competition between accessible hypersurface branches.
		
	\end{abstract}
	\tableofcontents
	
	\section{Introduction}
	The interior of a black hole can continue to evolve even after many observables of the boundary quantum system have approached equilibrium. For a two-sided AdS black hole, the Einstein--Rosen bridge provides a geometric example of this persistent evolution \cite{HartmanMaldacena,Susskind2014,StanfordSusskind2014}. Holographic complexity was proposed to describe aspects of this late-time behavior. The complexity=volume (CV) proposal relates complexity to the volume of a maximal bulk hypersurface \cite{Susskind2014,StanfordSusskind2014}, whereas the complexity=action (CA) proposal relates it to the gravitational action of the Wheeler--DeWitt patch \cite{Brown1,Brown2}. Subsequent work developed these prescriptions and explored their applications \cite{Couch2017,ChapmanPolicastro,Cai:2020wpc}. Previous calculations have examined how complexity responds to charged black branes, momentum relaxation, and black-hole thermodynamics \cite{BabaeiMomentum2020,BabaeiThermodynamic2022}. Complexity growth has also been investigated in geometries with Lifshitz scaling and hyperscaling violation using the CA prescription \cite{Alishahiha:2018tep}. The action of spacetime subregions has also been used to examine how different regions contribute to holographic complexity \cite{AlishahihaSubregion2019}. These studies motivate asking not only how fast complexity grows, but also which parts of the bulk geometry control that growth.
	
	Other approaches study complexity and information growth directly in quantum systems. Circuit complexity asks how many elementary operations are needed to prepare a target state from a chosen reference state, for a specified set of gates \cite{JeffersonMyers2017}. Krylov complexity tracks how a time-evolved operator spreads in a basis generated by repeated commutators with the Hamiltonian \cite{Parker2019}. Its behavior has been studied in Lifshitz-type scalar and Dirac field theories \cite{VasliKrylov2024,ImaniKrylov2025}. The related notion of spread complexity applies a Krylov construction to the evolution of quantum states \cite{Balasubramanian2022}. The growth rate of spread complexity has been related to the proper radial momentum of an infalling bulk probe in AdS$_3$ \cite{CaputaMomentum2024}. This proposal has also been explored in geometries with Lifshitz scaling and hyperscaling violation \cite{BitaghsirFadafan:2026lek}. Out-of-time-order correlators (OTOCs) provide a different diagnostic: they probe the growth of initially small perturbations and the scrambling of quantum information \cite{MSS2016}. Although these approaches address related questions about quantum dynamics, they define different quantities.
	
	The complexity-equals-anything (CEA) proposal provides a useful way to address this question \cite{Belin2022,Belin2023,MyersRuan2024}. Instead of extremizing the ordinary volume alone, it allows a local scalar to weight the contribution from each radial position. Different weights can therefore emphasize different features of the same black-hole interior. Curvature-dependent CEA observables have been used to probe black-hole interiors and singularities \cite{Jorstad2023,WangJiangLiu2023}, while matter-dependent observables constructed from the electromagnetic field strength have also been investigated \cite{JiangWangLiu2024}. When the boundary-time map has multiple accessible branches, distinct extremal hypersurfaces exist at the same boundary time. Comparing their on-shell values identifies the dominant hypersurface, as illustrated by studies of charged and Gauss--Bonnet AdS black holes \cite{WangJiangLiu2023,JiangWangLiu2024,WangLiWang2023}. These questions are especially interesting in charged geometries, where inner horizons and large-curvature regions can make the set of accessible hypersurfaces nontrivial \cite{JiangLiu2025}. We treat CEA as a family of gravitational complexity observables; different weights need not have the same boundary interpretation.
	
	Generalized Born--Infeld (GBI) electrodynamics provides a well-controlled setting for this investigation. Ordinary Born--Infeld theory modifies Maxwell electrodynamics at high field strength and also appears in effective descriptions of strings and D-branes \cite{BornInfeld,FradkinTseytlin1985,Tseytlin1996,Sorokin2022}. Its AdS black holes already have interesting charged horizon structures \cite{Dey2004,Cai2004}. Holographic complexity in Born--Infeld electrodynamic black holes has previously been studied using CA and CV 2.0 \cite{TaoWangYang2017,Meng2019}. Here we instead investigate weighted codimension-one CEA observables in GBI electrodynamics. The general framework of nonlinear electrodynamics was developed in Refs.~\cite{Plebanski1968,Boillat1970}. GBI extends Born--Infeld electrodynamics while preserving continuous electric--magnetic duality \cite{Bandos2020}. The Gaillard--Zumino self-duality condition \cite{GaillardZumino,GaillardZumino1997,GibbonsRasheed} and the Courant--Hilbert construction \cite{CourantHilbert,RussoTownsend2024} organize such self-dual nonlinear theories. Recent studies have connected nonlinear electrodynamics to stress-tensor deformations and classified causal self-dual theories \cite{Babaei2022,BabaeiRootTT2025,BabaeiRootUnify2025,BabaeiParity2026}. A related construction of causal NLED black-hole solutions is given in Ref.~\cite{BabaeiUnifiedBH2025}. Causality and energy conditions impose further constraints on physically admissible nonlinear-electromagnetic theories \cite{RussoTownsend2024,RussoTownsendEC2024}.
	
	There are three specific reasons to study complexity in the GBI geometry. First, the electric black-hole solution is known explicitly, so we can compare different complexity weights without changing the underlying metric. Second, the black holes admit spherical, planar, and hyperbolic horizons. We can therefore see how topology changes the interior geometry and the hypersurfaces that can reach the boundaries. Third, at pole cancellation the nonlinear electromagnetic contribution exactly cancels the leading $1/r$ term in the blackening function, while the curvature singularity remains. This gives a useful test of whether different weighted potentials respond to the same singular geometry in the same way. In the electric sector, the ModMax coupling enters the metric through the dressed charge; the charge and Born--Infeld deformation can then be varied separately within a fixed-horizon family.
	
	Our main goal is to identify what the matter source, the stress trace, and the Weyl curvature reveal about these interiors. We first examine their effective potentials and the turning points allowed by the radial equations. We then calculate the boundary times and complexity growth of the accessible hypersurfaces. Finally, we compare the sensitivity to charge and Born--Infeld deformation and derive the condition under which two hypersurface branches could exchange dominance.
	
	Section~\ref{sec:gbi} introduces the GBI solution, and Section~\ref{sec:geometry} examines its horizons and pole-cancellation condition. We derive the extremal-hypersurface equations in Section~\ref{sec:cea} and compare the three scalar weights in Section~\ref{sec:weights}. Sections~\ref{sec:potentials} and \ref{sec:complexity} present the turning-point and complexity results. Section~\ref{sec:conclusion} summarizes the conclusions and remaining questions.
	
	\section{Self-dual nonlinear electrodynamics and the GBI black hole}
	\label{sec:gbi}
	
	\subsection{Courant--Hilbert form of the matter theory}
	
	We consider four-dimensional Einstein gravity coupled to nonlinear electrodynamics with a negative cosmological constant $\Lambda=-3/\LAdS^2$, described by the action
	\begin{equation}
		I=\frac{1}{16\pi\GN}\int \dd^4x\,\sqrt{-g}
		\left[R-2\Lambda+4\mathcal L(\mathcal S,\mathcal P)\right].
		\label{eq:action}
	\end{equation}
	Here $\mathcal S=-F_{\mu\nu}F^{\mu\nu}/4$ and $\mathcal P=-F_{\mu\nu}\widetilde F^{\mu\nu}/4$, with $\widetilde F^{\mu\nu}=\epsilon^{\mu\nu\rho\sigma}F_{\rho\sigma}/(2\sqrt{-g})$. The excitation $G^{\mu\nu}=-2\partial\mathcal L/\partial F_{\mu\nu}$ obeys $\nabla_\mu G^{\mu\nu}=0$. Continuous electric-magnetic duality requires the Gaillard--Zumino self-duality condition, which constrains the full two-invariant Lagrangian even when the solution of interest is purely electric \cite{GaillardZumino,GaillardZumino1997,GibbonsRasheed}.
	
	In the Courant--Hilbert parametrization, the nonnegative variables $U$ and $V$ satisfy $V-U=\mathcal S$ and $4UV=\mathcal P^2$. Electromagnetic self-duality then requires $\mathcal L_U\mathcal L_V=-1$. In the parity-invariant sector, a generating function specifies the theory \cite{CourantHilbert,RussoTownsend2024}. Suitable monotonicity and convexity conditions select a causal branch \cite{RussoTownsend2024}; causality also constrains the energy conditions \cite{RussoTownsendEC2024}.
	
	The generalized Born--Infeld Lagrangian is given by \cite{Bandos2020,Babaei2022,BabaeiUnifiedBH2025}
	\begin{equation}
		\mathcal L_{\rm GBI}
		=\frac{1}{\lBI}
		\left[1-\sqrt{
			1-2\lBI\left(
			\mathcal S\cosh\gamma
			+\sqrt{\mathcal S^2+\mathcal P^2}\sinh\gamma
			\right)-\lBI^2\mathcal P^2}\right].
		\label{eq:GBIL}
	\end{equation}
	Here $\lBI>0$ denotes the Born--Infeld deformation parameter, and $\gamma\geq0$ is the dimensionless ModMax coupling. Eq.~\eqref{eq:GBIL} reduces to Born--Infeld theory at $\gamma=0$ and to ModMax as $\lBI\to0$ \cite{Bandos2020}. The Courant--Hilbert generating function for this theory, together with the conditions required for causality and a real-valued Lagrangian, is discussed in Refs.~\cite{BabaeiRootTT2025,RussoTownsend2024}. Below, we need only its electric sector and the resulting gravitational solution.
	
	\subsection{Electric topological AdS solution}
	
	Consider the static metric and electric potential \cite{BabaeiUnifiedBH2025}
	\begin{equation}
		\dd s^2=-f(r)\dd t^2+\frac{\dd r^2}{f(r)}
		+r^2\dd\Omega_{k,2}^2,
		\qquad A=A_t(r)\dd t,
		\label{eq:metric}
	\end{equation}
	where the unit horizon section has constant curvature $k=1,0,-1$ and volume $V_k$, understood to be regulated for noncompact horizons \cite{Birmingham1999}. In the purely electric sector $\mathcal P=0$, we use the dressed charge $\qg=\ee^{-\gamma/2}Q$. The electric field and the corresponding gauge potential are \cite{BabaeiUnifiedBH2025}
	\begin{equation}
		\begin{aligned}
			A_t'(r)&=-\frac{\ee^{-\gamma}Q}
			{\sqrt{r^4+\lBI\qg^2}},\\
			A_t(r)&=\frac{\ee^{-\gamma}Q}{r}\,
			{}_2F_1\!\left(\frac14,\frac12;\frac54;
			-\frac{\lBI\qg^2}{r^4}\right).
		\end{aligned}
		\label{eq:electricfield}
	\end{equation}
	Thus all metric functions depend on $Q$ and $\gamma$ only through $\qg$. The nonzero mixed components of the matter stress tensor are
	\begin{equation}
		T^t{}_t=T^r{}_r=
		\frac{1-\sqrt{1+\lBI\qg^2/r^4}}{\lBI},
		\qquad
		T^\theta{}_\theta=T^\phi{}_\phi=
		\frac{1-r^2/\sqrt{r^4+\lBI\qg^2}}{\lBI}.
		\label{eq:stresscomponents}
	\end{equation}
	The electric field stays finite as $r\to0$, but the energy density diverges as $-T^t{}_t\sim \qg/(\sqrt{\lBI}\,r^2)$. Even so, its radial integral $\int_0^\epsilon\dd r\,r^2(-T^t{}_t)$ is finite. This allows the electromagnetic contribution to cancel the $1/r$ term in $f(r)$. As we show below, the cancellation does not remove the curvature singularity.
	
	Substituting the stress tensor into Einstein's equations reduces the gravitational system to a radial equation for the blackening function,
	\begin{equation}
		r f'(r)+f(r)-k+\Lambda r^2
		+\frac{2}{\lBI}
		\left(\sqrt{r^4+\lBI\qg^2}-r^2\right)=0 .
		\label{eq:radialEinstein}
	\end{equation}
	The angular Einstein equation follows from Eq.~\eqref{eq:radialEinstein} by differentiation. Integrating this equation, we obtain \cite{BabaeiUnifiedBH2025}
	\begin{equation}
		\begin{split}
			f(r)={}&k-\frac{\omega_0}{r}+\frac{r^2}{\LAdS^2}
			+\frac{2r^2}{3\lBI}
			-\frac{2\sqrt{r^4+\lBI\qg^2}}{3\lBI}
			+\frac{4\qg^2}{3r^2}
			{}_2F_1\!\left(\frac14,\frac12;\frac54;
			-\frac{\lBI\qg^2}{r^4}\right),
		\end{split}
		\label{eq:blackening}
	\end{equation}
	where we have used \(\Lambda=-3/\LAdS^2\). Fixing the horizon at $r=\rh$ gives
	\begin{equation}
		\begin{split}
			\omega_0={}&k\rh+\frac{\rh^3}{\LAdS^2}
			+\frac{2\rh^3}{3\lBI}
			-\frac{2\rh\sqrt{\rh^4+\lBI\qg^2}}{3\lBI}
			+\frac{4\qg^2}{3\rh}
			{}_2F_1\!\left(\frac14,\frac12;\frac54;
			-\frac{\lBI\qg^2}{\rh^4}\right).
		\end{split}
		\label{eq:masshorizon}
	\end{equation}
	The Hawking temperature is
	\begin{equation}
		T=\frac{f'(\rh)}{4\pi}
		=\frac{k\lBI+(2+3\lBI/\LAdS^2)\rh^2
			-2\sqrt{\rh^4+\lBI\qg^2}}
		{4\pi\lBI\rh}.
		\label{eq:temperature}
	\end{equation}
	
	\section{Extremality, pole cancellation, and topology}
	\label{sec:geometry}
	
	An extremal horizon has zero temperature: $T(r_e)=0$. Solving Eq.~\eqref{eq:temperature} for $T(r_e)=0$ and choosing the positive-radius solution gives
	\begin{equation}
		r_e^2=\frac{\LAdS^2}{3(3\lBI+4\LAdS^2)}
		\left[
		2\sqrt{k^2\LAdS^4+3\qg^2(3\lBI+4\LAdS^2)}
		-k(3\lBI+2\LAdS^2)
		\right].
		\label{eq:extremalradius}
	\end{equation}
	The spherical branch requires $2\qg>\sqrt{\lBI}$. The planar branch exists for $\qg>0$, and the hyperbolic branch also admits $\qg=0$; the latter gives $r_e=\LAdS/\sqrt3$. The extremal mass parameter follows from Eq.~\eqref{eq:masshorizon} at this radius.
	
	For nonzero charge, the small-$r$ expansion of Eq.~\eqref{eq:blackening} gives
	
	\begin{equation}
		f(r)=\frac{A_{-1}}{r}
		+k-\frac{2\qg}{\sqrt{\lBI}}
		+\left(\frac{1}{\LAdS^2}+\frac{2}{3\lBI}\right)r^2
		-\frac{r^4}{5\lBI^{3/2}\qg}+O(r^8).
		\label{eq:smallrf}
	\end{equation}
	where
	\begin{equation}
		\begin{aligned}
			A_{-1}&=-\omega_0+
			\frac{\Gamma(\tfrac14)^2}{3\sqrt\pi}
			\frac{\qg^{3/2}}{\lBI^{1/4}},\\
			B_0&=k-\frac{2\qg}{\sqrt{\lBI}}.
		\end{aligned}
		\label{eq:Aminusone}
	\end{equation}
	The $1/r$ term vanishes when
	\begin{equation}
		\omega_0=
		\frac{\Gamma(\tfrac14)^2}{3\sqrt\pi}
		\frac{\qg^{3/2}}{\lBI^{1/4}}
		\label{eq:polecancellation}
	\end{equation}
	
	\begin{figure}[!htbp]
		\centering
		\includegraphics[width=0.5\textwidth]{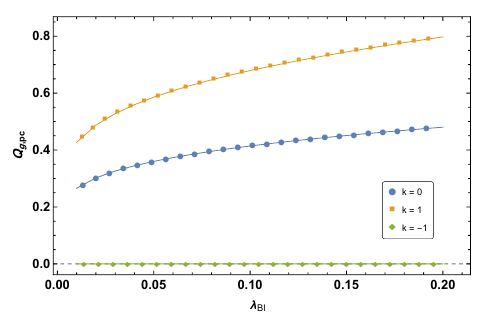}
		\caption{Pole-cancellation loci. At
			$\lBI=0.1$, the representative cancellation charges are
			$\qg_{\rm pc}\simeq0.4141$ for $k=0$ and
			$\qg_{\rm pc}\simeq0.6791$ for $k=1$. These marked points are
			used as reference values in the scans below; the neutral
			hyperbolic solution lies on the degenerate cancellation locus.}
		\label{fig:poleCancellation}
	\end{figure}

	The coefficient $A_{-1}$ contains the mass term and the nonlinear electromagnetic contribution. Their cancellation, Eq.~\eqref{eq:polecancellation}, defines the locus shown in Figure~\ref{fig:poleCancellation}. We use the marked planar and spherical values at $\lBI=0.1$ as reference points for later scans. With $\rh=\LAdS=1$, the neutral value is $A_{-1}=-(k+1)$. The spherical black hole therefore needs more charge to reach cancellation than the planar one. The neutral hyperbolic case instead has $A_{-1}=0$ and is regular pure AdS. 
	
	For charged, spherically symmetric Einstein--NLED solutions asymptotic to Reissner--Nordstr\"om spacetime, Russo and Townsend established that causal NLED leads to a central metric singularity and at most two Killing horizons \cite{RussoTownsendBH2026}. Related analyses of causal NLED black holes and the excision of Cauchy horizons appear in Refs.~\cite{BabaeiUnifiedBH2025,HaleHennigarKubiznak2026}. Single-horizon solutions and magnetic examples with finite electromagnetic self-energy have also been investigated \cite{GuzmanFathiVillanueva2026,FathiGuzmanVillanueva2026}. For the GBI-AdS geometries studied here, including non-spherical horizons, we analyze the singularity directly using the near-origin expansion. To see why charged pole cancellation does not render the black hole regular, consider the Kretschmann scalar of the metric
	\eqref{eq:metric}:
	\begin{equation}
		R_{\mu\nu\rho\sigma}R^{\mu\nu\rho\sigma}
		=
		f''(r)^2
		+\frac{4f'(r)^2}{r^2}
		+\frac{4\left[f(r)-k\right]^2}{r^4}.
		\label{eq:Kretschmann}
	\end{equation}
	Hence,
	\begin{equation}
		\begin{aligned}
			A_{-1}\neq 0:\qquad
			R_{\mu\nu\rho\sigma}R^{\mu\nu\rho\sigma}
			&=
			\frac{12A_{-1}^2}{r^6}
			+O(r^{-5}),
			\\
			A_{-1}=0,\quad \qg>0:\qquad
			R_{\mu\nu\rho\sigma}R^{\mu\nu\rho\sigma}
			&=
			\frac{16\qg^2}{\lBI\,r^4}
			+O(r^{-2}).
		\end{aligned}
		\label{eq:KretschmannLimits}
	\end{equation}
	Pole cancellation weakens the leading curvature divergence from $r^{-6}$ to $r^{-4}$, but the singularity remains. The neutral hyperbolic solution is different: $\omega_0=0$ and $f(r)=-1+r^2/\LAdS^2$. It describes regular pure AdS in hyperbolic slicing.

	We next ask how charge and Born--Infeld deformation change the geometry at a fixed outer-horizon radius. Using $f(1)=0$, the radial Einstein equation \eqref{eq:radialEinstein} gives
	\begin{equation}
		f(r)=f_0(r)+\frac{2}{\lBI r}\int_r^1
		\left(\sqrt{s^4+\lBI\qg^2}-s^2\right)\dd s,
		\label{eq:fInt}
	\end{equation}
	where $f_0(r)=k+r^2/\LAdS^2-(k+1/\LAdS^2)/r$ is the neutral solution with the same horizon radius. For $0<r<1$, $\qg>0$, and $\lBI>0$, differentiation parameters gives
	\begin{align}
		\partial_{\qg}f(r)
		&=\frac{2\qg}{r}\int_r^1
		\frac{\dd s}{\sqrt{s^4+\lBI\qg^2}}>0,
		\label{eq:dqf}\\
		\partial_{\lBI}f(r)
		&=-\frac{1}{\lBI^2 r}\int_r^1
		\frac{\bigl(\sqrt{s^4+\lBI\qg^2}-s^2\bigr)^2}
		{\sqrt{s^4+\lBI\qg^2}}\dd s<0.
		\label{eq:dlf}
	\end{align}
	Hence increasing the dressed charge raises $f(r)$ at a fixed interior radius, whereas increasing the Born--Infeld deformation lowers it. These relations also permit a comparison of the responses to equal infinitesimal fractional variations of the two parameters. At fixed $r$ and fixed outer-horizon radius, the exact derivatives above imply
	\begin{equation}
		0<\frac{|\lBI\partial_{\lBI}f(r)|}
		{\qg\partial_{\qg}f(r)}
		\leq \frac{\lBI\qg^2}{8r^4}
		\leq \frac{\lBI\qg^2}{8r_0^4},
		\qquad r_0\leq r<\rh,
		\label{eq:geometricResponseBound}
	\end{equation}
	where $r_0>0$ is an arbitrary fixed lower radius. If $\lBI\qg^2/(8r_0^4)<1$, the blackening function responds less strongly to a fractional variation of the Born--Infeld deformation than to an equal fractional variation of the dressed charge throughout the interval. The value one is only a convenient comparison threshold; it does not signal a physical transition. For instance, with the representative planar parameters $\lBI=0.1$ and $\qg\simeq0.41407$, taking $r_0=0.8$ gives $\lBI\qg^2/(8r_0^4)\simeq5.23\times10^{-3}$.
	
	\begin{figure}[!htbp]
		\centering
		\includegraphics[width=\linewidth]{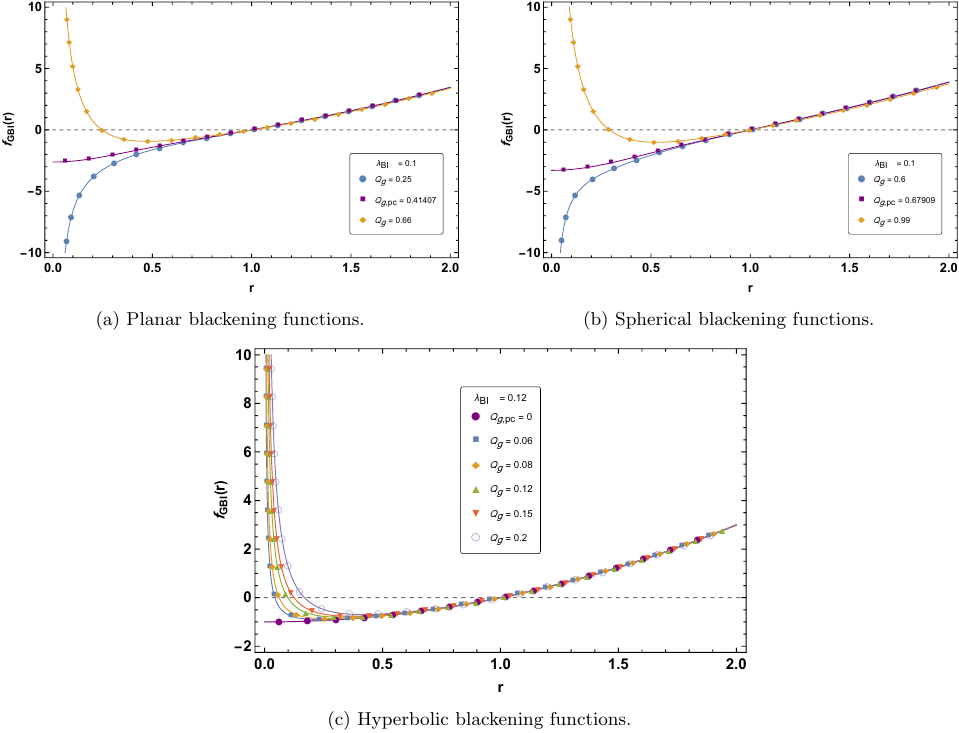}
		\caption{Representative GBI blackening functions. The planar and spherical charge scans pass through
			the pole-cancellation values marked in
			Figure~\ref{fig:poleCancellation}. For nonextremal horizons,
			$A_{-1}>0$ forces an inner horizon, while the neutral hyperbolic
			point is pure AdS in hyperbolic slicing.}
		\label{fig:blackeningGeometry}
	\end{figure}
	
	Figure~\ref{fig:blackeningGeometry} illustrates these derivative signs: at a fixed interior radius, increasing charge raises $f(r)$, while increasing Born--Infeld deformation lowers it. The coefficient $A_{-1}$ controls a different feature, namely the behavior near $r=0$. Just inside a nonextremal outer horizon, $f<0$. If $A_{-1}>0$, then $f$ becomes positive as $r\to0^+$, so there must be at least one inner horizon. If $A_{-1}<0$, this argument does not require an inner horizon. In the planar fixed-charge scan, $\lBI=0.1$ is the pole-cancellation point: decreasing $\lBI$ gives $A_{-1}>0$, while increasing it gives $A_{-1}<0$. The charged hyperbolic examples below have $A_{-1}>0$ and an inner horizon.
	
	\begin{figure}[!htbp]
		\centering
		\includegraphics[width=\linewidth]{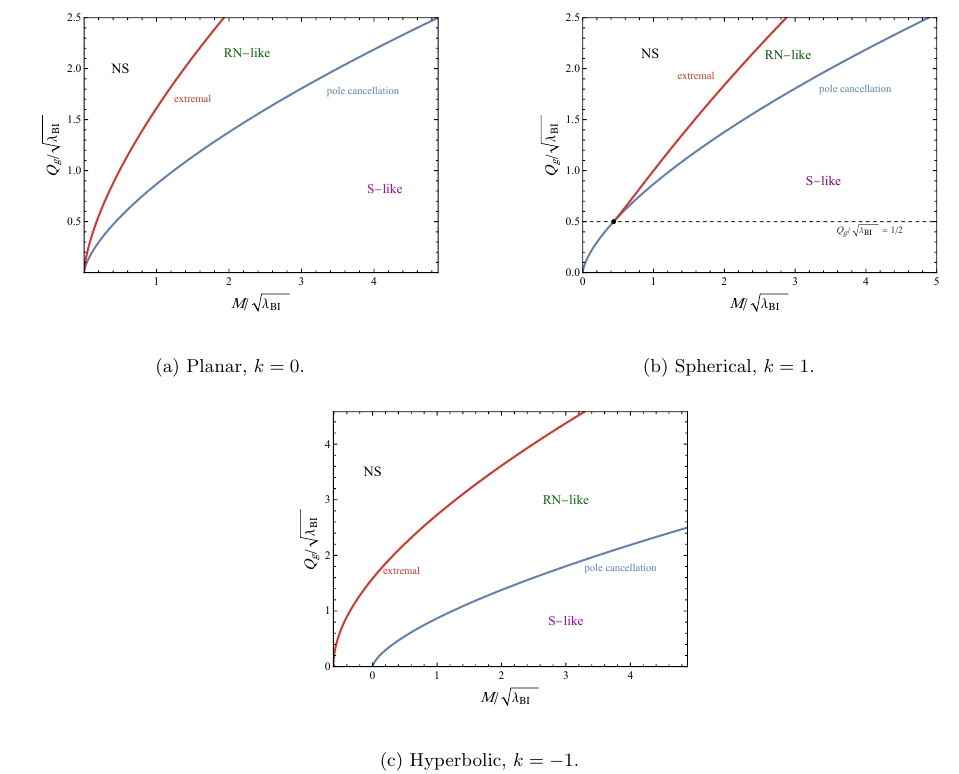}
		\caption{Mass--charge parameter spaces at $\lBI=0.1$ and $\gamma=0$. The axes use $M/\sqrt{\lBI}$
			and $Q_g/\sqrt{\lBI}$, with $M=\omega_0/2$ and
			$Q_g=\qg$. Red marks positive-radius extremality
			and blue marks $A_{-1}=0$. The labels NS, RN-like, and
			S-like identify horizonless, two-horizon, and one-horizon
			regions. In (b), the dashed charge threshold and black dot
			mark $\qg/\sqrt{\lBI}=1/2$ and the zero-radius meeting
			point of the curves. In (c), the blue curve ends at the
			plotting cutoff $\qg/\sqrt{\lBI}=2.5$, not at a physical
			endpoint.}
		\label{fig:massChargeDomains}
	\end{figure}
	
	Figure~\ref{fig:massChargeDomains} illustrates the regions associated with different horizon structures and the pole-cancellation condition in the mass--charge parameter space. Here $M=\omega_0/2$ is the mass parameter, so that the corresponding term in the blackening function is $-2M/r$, and $\qg$ denotes the dressed charge. At fixed $\lBI=0.1$, the pole-cancellation curve follows from Eq.~\eqref{eq:Aminusone}
	\begin{equation}
		\frac{M_{\rm pc}}{\sqrt{\lBI}}
		=
		\frac{\Gamma(\tfrac14)^2}{6\sqrt{\pi}}
		\left(
		\frac{\qg}{\sqrt{\lBI}}
		\right)^{3/2}.
		\label{eq:massChargePC}
	\end{equation}
	When mass and charge are varied independently, this curve does not depend on $k$. In contrast, the red extremal boundary, defined by $T(r_e)=0$ and $M=\omega_0(r_e,\qg)/2$, depends on the horizon topology. Where a positive-radius extremal solution exists, configurations with $M<M_{\rm ext}(\qg)$ contain naked singularities, while those satisfying $M_{\rm ext}(\qg)<M<M_{\rm pc}(\qg)$ have two horizons. For $M>M_{\rm pc}(\qg)$, the solutions have one horizon. The spherical extremal branch begins at $\qg/\sqrt{\lBI}=1/2$, whereas the hyperbolic extremal branch extends to $\qg=0$ at negative mass.
	
	\section{Weighted extremal hypersurfaces}
	\label{sec:cea}
	
	\subsection{One-dimensional reduction}
	
	Introduce the ingoing Eddington--Finkelstein coordinate $v=t+r_*(r)$, where $r_*'=1/f$.  The metric becomes
	\begin{equation}
		\dd s^2=-f(r)\dd v^2+2\dd v\dd r+r^2\dd\Omega_{k,2}^2.
		\label{eq:EFmetric}
	\end{equation}
	For a two-sided geometry with equal boundary times $t_L=t_R=\tau/2$, we follow the codimension-one CEA construction \cite{Belin2022,Belin2023,Jorstad2023} and consider the functional
	\begin{equation}
		\mathcal C_X(\tau)
		=
		\frac{V_k}{\GN\LAC}
		\int_{\Sigma}\dd\sigma\,
		\left(\frac{r}{\LAC}\right)^2
		\aX(r)
		\sqrt{-f(r)\dot v^2+2\dot v\dot r},
		\label{eq:CEAfunctional}
	\end{equation}
	to be extremized over hypersurfaces $\Sigma$ anchored at the prescribed boundary times. The scale $\LAC$ fixes the normalization of the observable, and $\aX(r)$ is a dimensionless local weight. For a local scalar $X(r)$ of length dimension $-\Delta_X$, we define the weight as
	\begin{equation}
		\aX(r)
		=
		1+\alpha_{\rm AC}\LAC^{\Delta_X}X(r),
		\label{eq:weightgeneric}
	\end{equation}
	where $\alpha_{\rm AC}$ denotes the CEA coupling. Since the three scalars considered below are nonnegative, we choose nonnegative CEA couplings to ensure positive weights. The same weight is used on both halves of the symmetric hypersurface. The volume proposal is recovered for $\aX(r)=1$. We use reparametrization invariance to choose the parameter along the hypersurface such that
	\begin{equation}
		\sqrt{-f\dot v^2+2\dot v\dot r}
		=\aX(r)\left(\frac{r}{\LAC}\right)^2.
		\label{eq:gauge}
	\end{equation}
	This gauge simplifies the equations of motion. Since the Lagrangian does not depend explicitly on $v$, the Euler--Lagrange equation implies that the conjugate momentum $P_v$ is conserved. In the chosen gauge, it takes the form $\PV=\dot r-f\dot v$. Combining this momentum with the gauge condition gives the radial equation
	\begin{equation}
		\dot r^2+\Veff(r)=\PV^2,
		\qquad
		\Veff(r)=-f(r)\aX(r)^2\left(\frac{r}{\LAC}\right)^4.
		\label{eq:Ueff}
	\end{equation}
	The motion of the hypersurface can therefore be described in terms of an effective potential $\Veff(r)$. At its turning point, the radial velocity vanishes, $\dot r=0$, which gives
	\begin{equation}
		\PV^2=\Veff(\rmin).
		\label{eq:turningpoint}
	\end{equation}
	However, not every solution of this equation corresponds to a hypersurface that can reach the AdS boundary. The radial motion must remain real along the entire outward trajectory. This requires
	\begin{equation}
		\PV^2-\Veff(r)\geq0
		\qquad \text{for every } r\geq\rmin,
		\label{eq:boundaryAccessibility}
	\end{equation}
	For every allowed turning point, the conserved momentum fixes the boundary time through \cite{Belin2022,Jorstad2023,WangJiangLiu2023,Carmi2017}
	\begin{equation}
		\frac{\tau}{\LAdS}=
		\left|-\frac{2}{\LAdS}
		\operatorname{PV}\!\int_{\rmin}^{\infty}
		\frac{\PV\,\dd r}{f(r)\sqrt{\PV^2-\Veff(r)}}\right|.
		\label{eq:timeintegral}
	\end{equation}
	We evaluate the simple-horizon poles in Eq.~\eqref{eq:timeintegral} using the Cauchy principal value. The square-root singularity at an ordinary turning point is integrable. The boundary time is therefore finite away from a critical momentum.
	
	At a boundary-accessible local maximum $r_f$ of $\Veff$,
	\begin{equation}
		\Veff'(r_f)=0,
		\qquad
		\Veff''(r_f)<0,
		\qquad
		\PV^2\longrightarrow P_\infty^2\equiv\Veff(r_f),
		\label{eq:latepeak}
	\end{equation}
	the double zero of $\PV^2-\Veff$ yields a logarithmic divergence in Eq.~\eqref{eq:timeintegral} as $\PV\to P_\infty$. A deep branch can approach the same critical momentum from above if its outward trajectory remains accessible. A vertical asymptote in a boundary-time plot then marks a critical-momentum limit, while ordinary turning points have finite boundary times \cite{Belin2022,Jorstad2023}.

	\subsection{Renormalized observable and competing branches}
	
	For a fixed background and scalar weight, the hypersurface functional diverges at the AdS boundary. We subtract the value of the symmetric surface at $\tau=0$ to remove this common divergence. We also choose the normalization so that the Hamilton--Jacobi relation has a simple form:
	\begin{equation}
		\Delta C(\tau)=\frac{\GN\LAC}{V_k\LAdS}
		\left[\mathcal C_X(\tau)-\mathcal C_X(0)\right].
		\label{eq:complexity}
	\end{equation}
	At a radial cutoff $R$, a numerically stable expression is
	\begin{equation}
		\Delta C=
		\frac{2}{\LAdS}\left[
		\int_{\rmin}^{\rh}\!\mathcal I_{\PV}(r)\dd r
		+\int_{\rh}^{R}\!\left(\mathcal I_{\PV}(r)-\mathcal I_0(r)\right)\dd r
		\right],
		\label{eq:subtractedC}
	\end{equation}
	where
	\begin{equation*}
		\mathcal I_{\PV}(r)=
		\frac{\aX(r)^2(r/\LAC)^4}
		{\sqrt{\PV^2-\Veff(r)}},
		\qquad
		\mathcal I_0(r)=
		\frac{\aX(r)^2(r/\LAC)^4}
		{\sqrt{-\Veff(r)}}.
	\end{equation*}
	For each smooth extremal-hypersurface branch, the Hamilton--Jacobi relation \cite{Belin2022,Belin2023,WangJiangLiu2023} gives
	\begin{equation}
		\frac{\dd\Delta C^{(i)}}{\dd x}=\PV^{(i)}(x),
		\qquad x\equiv\frac{\tau}{\LAdS}.
		\label{eq:HJ}
	\end{equation}
	This equation holds on each smooth on-shell branch. If two branches exchange dominance with different momenta at the crossing time, the complexity has a kink at that time.
	
	Several extremal hypersurfaces can exist at the same boundary time. The physical prescription is to compare their on-shell complexities and select the largest value among the boundary-accessible branches \cite{Belin2023,Jorstad2023,WangJiangLiu2023,JiangWangLiu2024},
	\begin{equation}
		\Delta C(\tau)=\max_i
		\Delta C^{(i)}(\tau).
		\label{eq:envelope}
	\end{equation}
	When $\tau(\PV)$ is not one-to-one, several turning points can correspond to the same boundary time. Eq.~\eqref{eq:envelope} then selects the largest complexity among the accessible branches.

	\section{Matter, stress-trace, and Weyl weights}
	\label{sec:weights}
	
	The near-singularity behavior of a weighted effective potential can be classified according to the leading radial powers of the blackening function and scalar weight. Consider the general asymptotic forms
	\begin{equation}
		f(r)\sim F_p r^{-p},
		\qquad
		a_X(r)\sim b_X r^{-s_X},
		\qquad r\to0^+,
		\label{eq:generalScaling}
	\end{equation}
	where $F_p$ and $b_X$ are nonzero coefficients, with $p\geq0$ and $s_X\geq0$. In particular, a finite nonzero scalar weight corresponds to $s_X=0$. For a scalar behaving as $X(r)\sim c_Xr^{-s_X}$ with $s_X>0$, the weight defined in Eq.~\eqref{eq:weightgeneric} has the same leading exponent when its coupling is nonzero.
	
	Substituting Eq.~\eqref{eq:generalScaling} into the effective potential \eqref{eq:Ueff} gives
	\begin{equation}
		U_X(r)\sim
		-\frac{F_p b_X^2}{L_A^4}
		r^{\,4-p-2s_X}.
		\label{eq:generalPotentialScaling}
	\end{equation}
	The near-singularity behavior of the effective potential is therefore determined by the exponent $4-p-2s_X$. When $F_p<0$, corresponding to a positive effective potential near the singularity, three distinct classes follow:
	\begin{equation}
		\begin{aligned}
			s_X<\frac{4-p}{2}
			&:\quad U_X(r)\longrightarrow0,\\
			s_X=\frac{4-p}{2}
			&:\quad U_X(r)\longrightarrow U_{X,0}>0,\\
			s_X>\frac{4-p}{2}
			&:\quad U_X(r)\longrightarrow+\infty.
		\end{aligned}
		\label{eq:potentialClassification}
	\end{equation}
	For $F_p>0$, the same exponents govern the potential, but its leading sign is negative.
	
	This classification explains why our three scalar weights probe the same singularity differently. At charged pole cancellation, $f(r)\to B_0<0$ and hence $p=0$. The dividing value is $s_X=2$. We choose scalar weights with $s_X=0$, $s_X=2$, and $s_X>2$. Their effective potentials then tend to zero, a finite positive limit, and infinity, respectively. We now calculate these limits explicitly.
	
	\subsection{Three scalar prescriptions}
	
	Electromagnetic field-strength weights have previously been used in generalized volume-complexity \cite{JiangWangLiu2024}. Here we extend this choice to the on-shell GBI matter Lagrangian,
	\begin{equation}
		\mathcal L_{\rm m}(r)=\frac{1}{\lBI}
		\left(1-\frac{r^2}{\sqrt{r^4+\lBI\qg^2}}\right).
		\label{eq:Lmatter}
	\end{equation}
	For $\qg>0$, this scalar is positive. It approaches $1/\lBI$ near $r=0$ and decays as $\qg^2/(2r^4)$ near the AdS boundary. We set $X=\mathcal L_{\rm m}$ and $\Delta_X=2$ in Eq.~\eqref{eq:weightgeneric}, with coupling $\alpha_{\rm m}$. Its weight stays finite at the singularity, so $s_X=0$. The associated potential therefore vanishes at the pole-cancellation endpoint.
	
	The second scalar is the negative trace of the nonlinear-electromagnetic stress tensor,
	\begin{equation}
		X_T(r)\equiv-T^\mu{}_{\mu}
		=\frac{2}{\lBI}\left[
		\sqrt{1+\frac{\lBI\qg^2}{r^4}}
		+\left(1+\frac{\lBI\qg^2}{r^4}\right)^{-1/2}
		-2
		\right].
		\label{eq:stressTrace}
	\end{equation}
	Here we have used the on-shell components in Eq.~\eqref{eq:stresscomponents}. The stress tensor of a self-dual nonlinear electromagnetic theory is invariant under electromagnetic duality rotations \cite{GaillardZumino,FerkoStress2023}, and hence so is its trace. It therefore provides a natural matter-based probe of the duality-invariant sector. Unlike the on-shell Lagrangian, $X_T$ diverges as $r^{-2}$ near the charged singularity. We take $\Delta_X=2$ and write its coupling as $\alpha_T$. For nonzero coupling, the resulting weight has $s_X=2$, corresponding to a finite nonzero effective-potential limit at charged pole cancellation. For the charged pole-cancellation backgrounds in our scans, the finite endpoint is a one-sided local maximum.
	
	Finally, following previous curvature-dependent CEA prescriptions \cite{Belin2022,Belin2023,Jorstad2023,WangJiangLiu2023}, we consider the square of the Weyl tensor. For the metric \eqref{eq:metric}, it takes the form
	\begin{equation}
		C^2\equiv C_{\mu\nu\rho\sigma}C^{\mu\nu\rho\sigma}
		=\frac{4}{3r^8}\Bigg[
		3\omega_0r
		-\frac{2\qg^2r^2}{\sqrt{r^4+\lBI\qg^2}}\\
		-4\qg^2\,{}_2F_1\!\left(
		\frac14,\frac12;\frac54;
		-\frac{\lBI\qg^2}{r^4}
		\right)\Bigg]^2.
		\label{eq:Cgeneral}
	\end{equation}
	The Weyl coupling $\alpha_W$ is defined through $X=C^2$ and $\Delta_X=4$. This scalar characterizes tidal curvature, in contrast to the two matter-based prescriptions. At charged pole cancellation, $C^2$ diverges as $r^{-4}$, so the corresponding weight has $s_X=4$ for nonzero coupling. It therefore belongs to the class of divergent positive effective potentials. Away from pole cancellation, the leading Weyl divergence is stronger, with $C^2\sim r^{-6}$. In the neutral limit, $C^2=12\omega_0^2/r^6$. All three weights approach unity as $r\to\infty$.
	
	The factorized potential \eqref{eq:Ueff} immediately implies, for $r>0$ and a positive weight,
	\begin{equation}
		\operatorname{sgn} U_X(r)
		=-\operatorname{sgn} f(r).
		\label{eq:potSign}
	\end{equation}
	Thus a positive scalar weight can change the height or position of a potential maximum, but it cannot change the horizon radii. It also cannot introduce a new sign-changing zero of $U_X$ away from a zero of $f$. The scalar weight changes how the hypersurface probes the geometry, not the geometry itself.

	\subsection{The singularity hierarchy}
	
	Pole cancellation allows us to compare the three weights in the same geometry. Their behavior near $r=0$ is different even though $f(r)$ is unchanged. Using $A=A_{-1}$ and $B=B_0$ from Section~\ref{sec:geometry}, the charged matter scalars have the expansions
	\begin{align}
		\mathcal L_{\rm m}(r)&=\frac1{\lBI}
		-\frac{r^2}{\lBI^{3/2}\qg}+O(r^6),
		\label{eq:matterNearOrigin}\\
		X_T(r)&=\frac{2\qg}{\sqrt{\lBI}\,r^2}
		-\frac4{\lBI}+\frac{3r^2}{\lBI^{3/2}\qg}+O(r^6).
		\label{eq:traceNearOrigin}
	\end{align}
	The $r^2$ term in $f(r)$ drops out of the Weyl combination in Eq.~\eqref{eq:Cgeneral}. In particular,
	\begin{equation}
		\begin{aligned}
			A\ne0:\quad C^2(r)&=\frac{12A^2}{r^6}
			-\frac{16A\qg}{\sqrt{\lBI}\,r^5}
			+\frac{16\qg^2}{3\lBI\,r^4}+O(r^{-1}),
			\\
			A=0:\quad C^2(r)&=\frac{16\qg^2}{3\lBI\,r^4}
			+\frac{16}{5\lBI^2}+O(r^4).
		\end{aligned}
		\label{eq:WeylNearOriginLimits}
	\end{equation}
	Thus, cancellation of the geometric $1/r$ pole changes the leading Weyl divergence but leaves a curvature singularity.
	
	For positive CEA couplings, the resulting weighted potentials have the following leading behavior away from pole cancellation:
	\begin{equation}
		\begin{aligned}
			U_{\rm m}(r)&=-\frac{A}{\LAC^4}
			\left(1+\frac{\alpha_{\rm m}\LAC^2}{\lBI}\right)^2 r^3
			+O(r^4),
			\\
			U_T(r)&=-\frac{4\alpha_T^2\qg^2 A}{\lBI}\,\frac1r
			+O(1),
			\\
			U_W(r)&=-\frac{144\alpha_W^2\LAC^4 A^5}{r^9}
			+O(r^{-8}).
		\end{aligned}
		\label{eq:potentialsNearOrigin}
	\end{equation}
	At charged pole cancellation, $A=0$, these become
	\begin{equation}
		\begin{aligned}
			U_{\rm m}(r)&=-\frac{B}{\LAC^4}
			\left(1+\frac{\alpha_{\rm m}\LAC^2}{\lBI}\right)^2r^4+O(r^6),
			\\
			U_T(r)&=-\frac{4\alpha_T^2\qg^2B}{\lBI}+O(r^2),
			\\
			U_W(r)&=-\frac{256\alpha_W^2\LAC^4 B\qg^4}
			{9\lBI^2r^4}+O(r^{-2}).
		\end{aligned}
		\label{eq:potentialsNearOriginPC}
	\end{equation}
	For a charged pole-cancelled black hole with a positive-radius outer horizon, one has $B<0$. Therefore the leading potentials are positive toward the singularity, but their limits are qualitatively different,
	\begin{equation}
		U_{\rm m}\longrightarrow0,
		\qquad U_T\longrightarrow P_0^2>0,
		\qquad U_W\longrightarrow+\infty,
		\quad (r\to0^+,\ A=0),
		\label{eq:threeLimits}
	\end{equation}
	where
	\begin{equation}
		P_0=\frac{2\alpha_T\qg}{\sqrt{\lBI}}
		\sqrt{\frac{2\qg}{\sqrt{\lBI}}-k}.
		\label{eq:traceEndpointMomentum}
	\end{equation}
	To establish the nature of this endpoint, we retain the next term in the stress-trace potential. At pole cancellation, Eqs.~\eqref{eq:smallrf} and \eqref{eq:traceNearOrigin} give $f(r)=B_0+d_2r^2+O(r^4)$ and $a_T(r)=a_{-2}r^{-2}+a_0+O(r^2)$, where
	\begin{equation*}
		\begin{aligned}
			d_2&=\frac{1}{\LAdS^2}+\frac{2}{3\lBI},
			& a_{-2}&=\frac{2\alpha_T\LAC^2\qg}{\sqrt{\lBI}},
			& a_0&=1-\frac{4\alpha_T\LAC^2}{\lBI}.
		\end{aligned}
	\end{equation*}
	Substitution into Eq.~\eqref{eq:Ueff} yields
	\begin{equation}
		U_T(r)=P_0^2-c_T r^2+O(r^4),\qquad
		c_T=\frac{2B_0a_{-2}a_0+d_2a_{-2}^2}{\LAC^4}.
		\label{eq:traceEndpointMaximumProof}
	\end{equation}
	The sign of $c_T$ determines the local behavior. For the charged pole-cancellation backgrounds considered in our scans, we set $\lBI=0.1$. For the planar geometry ($k=0$), with $\qg\simeq0.41407$ and $\alpha_T=0.2$, we obtain $c_T\simeq21.30556$. For the spherical geometry ($k=1$), with $\qg\simeq0.67909$ and $\alpha_T=0.1$, we obtain $c_T\simeq9.90510$. Since both coefficients are positive, $U_T(r)<P_0^2$ for all sufficiently small $r>0$, and $U_T'(r)=-2c_T r+O(r^3)\to0$ as $r\to0^+$. This establishes a one-sided local maximum at the singular endpoint for both backgrounds, with limiting momentum $P_0$. A hypersurface approaching this endpoint must also satisfy the outward accessibility condition~\eqref{eq:boundaryAccessibility}. Whether this branch dominates is determined by comparing its on-shell complexity with those of the other accessible branches at the same boundary time.
	
	\section{From effective potential to interior penetration}
	\label{sec:potentials}
	
	Not every turning point of the effective potential can be reached from the AdS boundary. Equations~\eqref{eq:Ueff}--\eqref{eq:boundaryAccessibility} require the radial trajectory to remain real along its entire outward path. A turning point hidden behind a higher outer barrier fails this condition. Consequently, finding several local maxima does not automatically mean that several physical hypersurface branches are available.
	
	\subsection{Changes of potential shape with the scalar prescription}
	
\begin{figure}[!htbp]
		\centering
		\includegraphics[width=\linewidth]{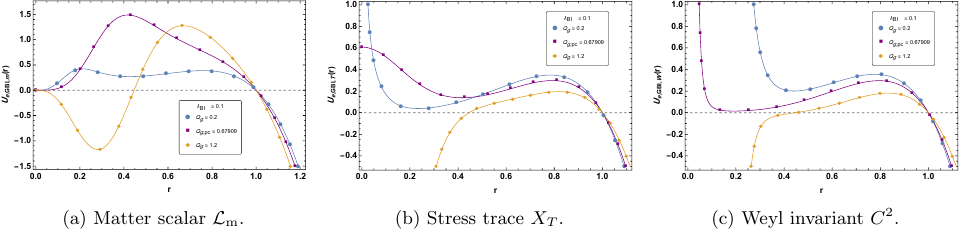}
		\caption{Spherical effective potentials at $\lBI=0.1$ for the charge scans of the three scalar prescriptions. The charge sets are those shown in the individual legends; the position and accessibility of positive barriers depend on both the background and the scalar weight.
			The CEA couplings are $(\alpha_{\rm m},\alpha_T,\alpha_W)=(1,0.1,5\times10^{-5})$.}
		\label{fig:UeffSpherical}
	\end{figure}

\begin{figure}[!htbp]
		\centering
		\includegraphics[width=\linewidth]{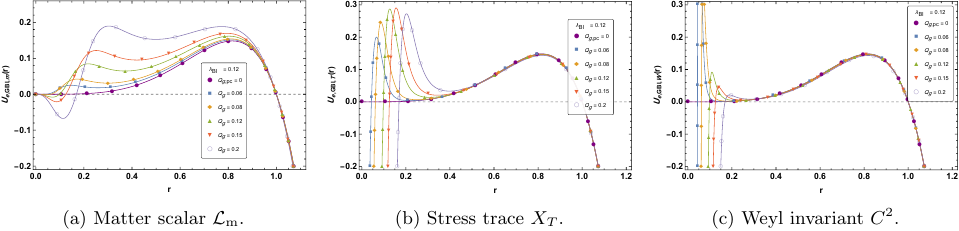}
		\caption{Hyperbolic effective potentials at $\lBI=0.12$. The neutral reference is locally AdS, while charged configurations exhibit nontrivial interior potentials.
			The CEA couplings are $(\alpha_{\rm m},\alpha_T,\alpha_W)=(3,3,5\times10^{-4})$.}
		\label{fig:UeffHyperbolic}
	\end{figure}

	Figures~\ref{fig:UeffSpherical} and \ref{fig:UeffHyperbolic} show the effective potentials for spherical and hyperbolic black holes. The three weights share the same horizon zeros because of Eq.~\eqref{eq:potSign}. Their shapes nevertheless differ inside the horizon. In particular, the limits in Eq.~\eqref{eq:threeLimits} explain why the matter, stress-trace, and Weyl potentials behave differently near pole cancellation.

	For a charged black hole with two horizons, $f>0$ inside the inner horizon. Since the weights are positive, $U_X<0$ there. A surface may pass through this region, but it cannot turn there with nonzero real momentum. Any deeper turning point must also pass the boundary-accessibility test \eqref{eq:boundaryAccessibility}.
	
	In the spherical scans of Figure~\ref{fig:UeffSpherical}, varying the charge across pole cancellation can change the inner-horizon structure. The three weights then give different positive potential barriers. In Figure~\ref{fig:UeffHyperbolic}, the neutral solution is a locally AdS reference, while the charged solutions have inner horizons and distinct deep-interior potentials.
	
	\subsection{Turning radii and boundary time}

	\begin{figure}[!htbp]
		\centering
		\includegraphics[width=\linewidth]{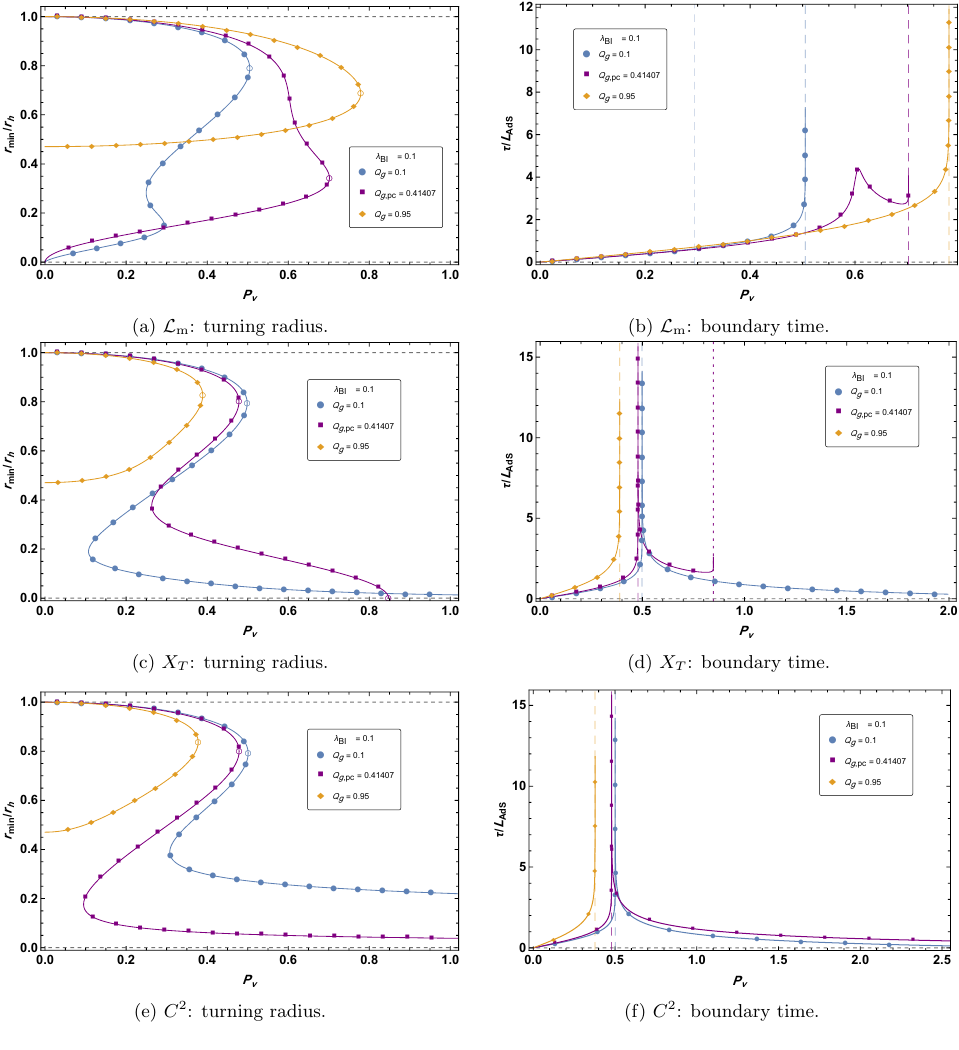}
		\caption{Planar turning-radius and boundary-time maps at $\lBI=0.1$ for matter, stress-trace, and Weyl weights. The nearly vertical time curves identify critical momentum limits.
			The CEA couplings are $(\alpha_{\rm m},\alpha_T,\alpha_W)=(1,0.2,10^{-4})$.}
		\label{fig:branchesPlanar}
	\end{figure}
	\begin{figure}[!htbp]
		\centering
		\includegraphics[width=\linewidth]{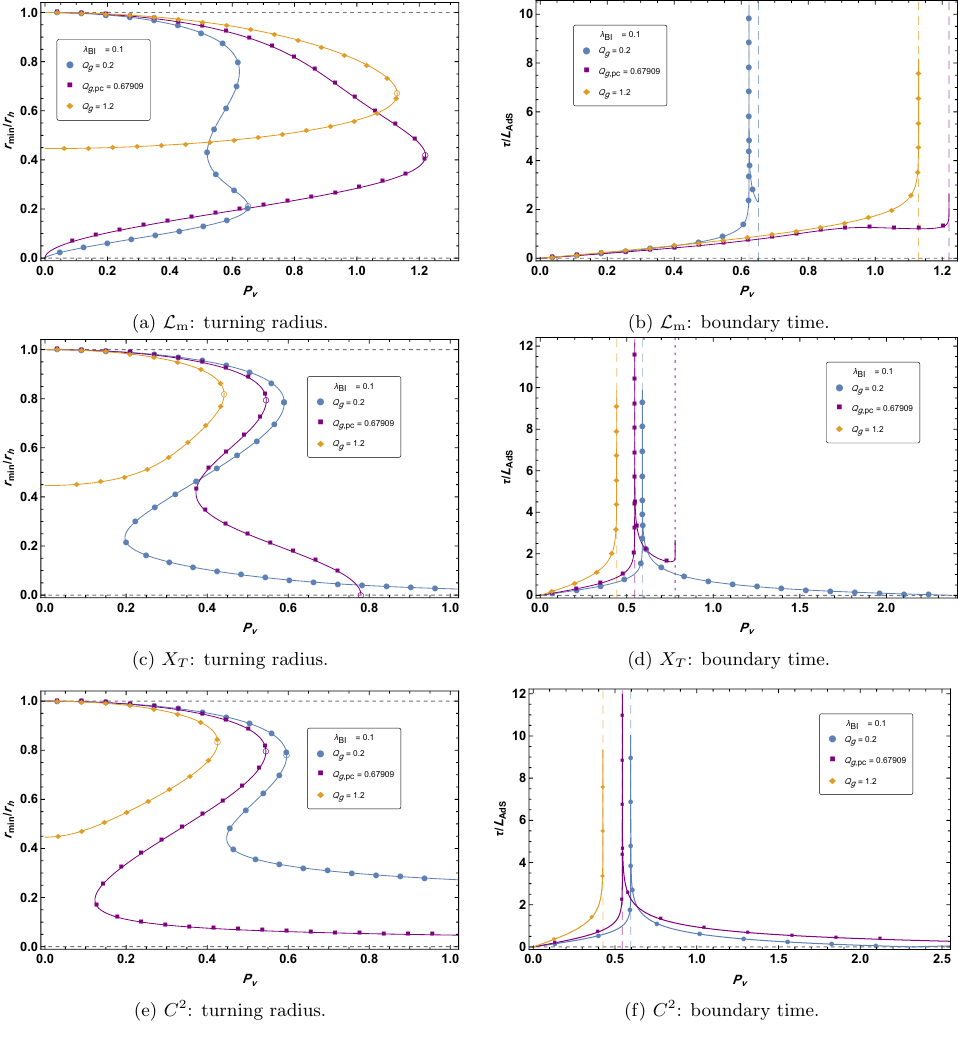}
		\caption{Spherical turning-radius and boundary-time maps at $\lBI=0.1$. The maps exhibit multiple candidate segments; outer barriers exclude deep turning roots that do not satisfy boundary accessibility.}
		\label{fig:branchesSpherical}
	\end{figure}
	\begin{figure}[!htbp]
		\centering
		\includegraphics[width=\linewidth]{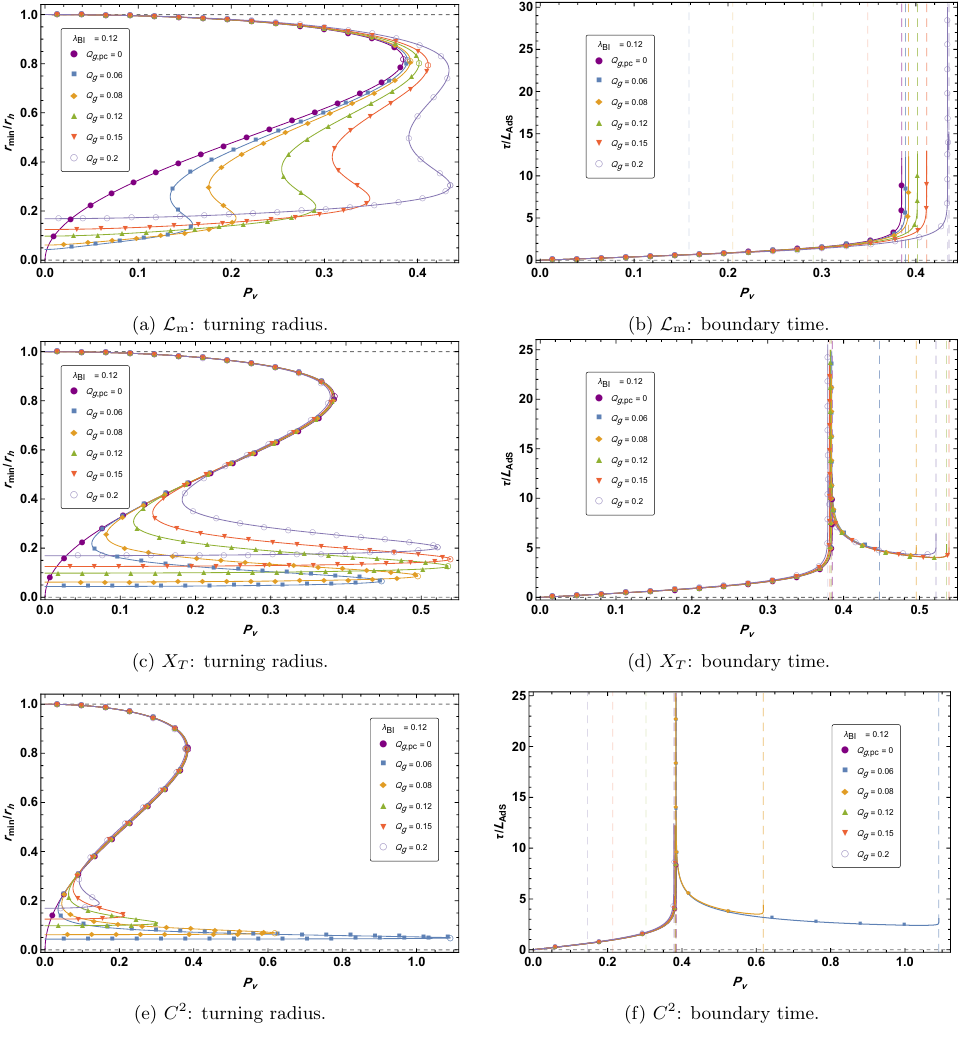}
		\caption{Hyperbolic turning-radius and boundary-time maps at $\lBI=0.12$. The neutral curve provides a reference for the charged turning-radius and time maps.}
		\label{fig:branchesHyperbolic}
	\end{figure}
	Equation~\eqref{eq:turningpoint} relates each turning radius to its conserved momentum, so that a local maximum of the positive effective potential corresponds to a local maximum of $P_v(r_{\min})$. The boundary time, however, depends on the entire radial trajectory through Eq.~\eqref{eq:timeintegral} and diverges when an accessible hypersurface approaches a nondegenerate critical double root. These features are illustrated in Figures~\ref{fig:branchesPlanar}--\ref{fig:branchesHyperbolic}, where the left panels show $r_{\min}$ as a function of $P_v$, and the right panels show the corresponding boundary time. The planar and spherical geometries exhibit multiple candidate branches, particularly near pole cancellation, while the hyperbolic scan also includes the neutral geometry. The folds in the turning-point maps reflect the extrema of the effective potential, whereas the vertical asymptotes in the boundary-time plots identify the corresponding accessible critical momenta.

	\subsection{Charge and Born--Infeld deformation as distinct parameters}
	\label{sec:parameterWeighted}
	
\begin{figure}[!htbp]
		\centering
		\includegraphics[width=\linewidth]{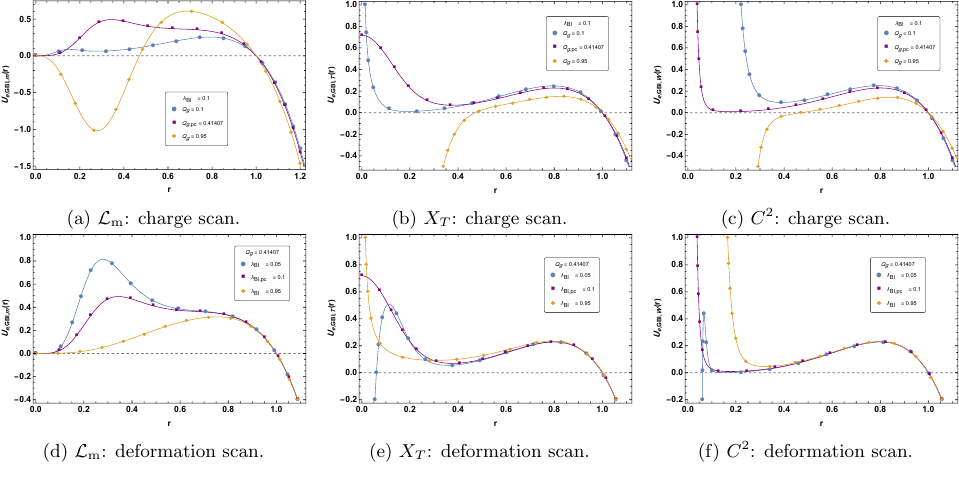}
		\caption{Planar effective potentials for the three scalar
			weights. The upper row varies the dressed charge at fixed
			$\lBI=0.1$; the lower row varies the Born--Infeld deformation at fixed
			$\qg=0.41407$, crossing the charged pole-cancellation
			configuration at $\lBI=0.1$. The CEA coupling for each
			scalar is held fixed within its corresponding scan.}
		\label{fig:lambdaFixedPotential}
	\end{figure}

	Having examined the structure of the effective potential, we now compare its response to variations in the dressed charge and the Born--Infeld deformation. Varying either parameter modifies both the blackening function and the local scalars entering the CEA weights. The resulting changes in the effective potentials are illustrated in Figure~\ref{fig:lambdaFixedPotential}. To understand the different responses to charge and deformation, we examine the geometric and scalar-weight contributions separately. For $\theta\in\{\qg,\lBI\}$, differentiation of $U_X(r)=U_0(r)a_X(r)^2$, where $U_0(r)=-f(r)(r/\LAC)^4$, gives the exact identity
	\begin{equation}
		\partial_\theta U_X(r)
		=-\left(\frac r{\LAC}\right)^4
		\left[a_X(r)^2\partial_\theta f(r)
		+2f(r)a_X(r)\partial_\theta a_X(r)\right],
		\qquad
		\partial_\theta a_X(r)
		=\alpha_X\LAC^{\Delta_X}\partial_\theta X(r).
		\label{eq:weightedParameterResponse}
	\end{equation}
	The first term describes the change of the geometry. The second describes the change of the scalar weight. For the unweighted potential, Eqs.~\eqref{eq:dqf} and \eqref{eq:dlf} imply that $\partial_{\qg}U_0(r)<0$ and $\partial_{\lBI}U_0(r)>0$ at a fixed interior radius $0<r<1$. For the full weighted potential, the scalar contribution either reinforces or opposes the geometric contribution, depending on the scalar and the background parameters.
	
	For the matter scalar, it is useful to write $S_r=\sqrt{r^4+\lBI\qg^2}$ and $\mathcal L_{\rm m}=\qg^2/[S_r(S_r+r^2)]$. Its derivatives at fixed radius are
	\begin{equation}
		\partial_{\qg}\mathcal L_{\rm m}
		=\frac{\qg r^2}{S_r^3}>0,\qquad
		\partial_{\lBI}\mathcal L_{\rm m}
		=-\frac{\qg^4(2S_r+r^2)}
		{2S_r^3(S_r+r^2)^2}<0.
		\label{eq:matterWeightResponse}
	\end{equation}
	For positive CEA coupling, increasing the charge enhances the matter weight, while increasing the Born--Infeld deformation suppresses it. Where $f(r)<0$, these changes oppose the corresponding geometric contributions. The response of the full matter potential is therefore determined by the interplay between the geometric and scalar-weight contributions. Both the charge and deformation scans in Figure~\ref{fig:lambdaFixedPotential}(a) and (d) cross the pole-cancellation point, resulting in a negative effective potential near the singularity on the $A_{-1}>0$ side.

	For the stress-trace weight, define $w=\sqrt{1+\lBI\qg^2/r^4}$. Its derivatives are
	\begin{equation}
		\begin{aligned}
			\partial_{\qg}X_T
			&=\frac{2\lBI\qg^3}{r^8w^3}>0,\\
			\partial_{\lBI}X_T
			&=\frac{(w-1)^2(1+2w-w^2)}
			{\lBI^2w^3}.
		\end{aligned}
		\label{eq:traceWeightResponse}
	\end{equation}
	For $\qg>0$, the charge derivative of $X_T$ is positive at all radii, whereas its deformation derivative changes sign at $r_*=\big[\lBI\qg^2/(2(1+\sqrt2))\big]^{1/4}$, being negative for $r<r_*$ and positive for $r>r_*$. In the region where $U_T>0$, increasing the charge reduces the potential through the geometric contribution but enhances it through the scalar-weight contribution. Increasing the Born--Infeld deformation, on the other hand, enhances the potential through the geometric contribution, while the scalar-weight contribution suppresses it for $r<r_*$ and enhances it for $r>r_*$. Thus, the two contributions compete for $r<r_*$ and reinforce each other for $r>r_*$ in the deformation scan. These competing and reinforcing effects govern the response of the stress-trace potential in panels (b) and (e) of Figure~\ref{fig:lambdaFixedPotential}.
	
	For the Weyl weight, define $\mathcal W(r)=r^2f''(r)-2rf'(r)+2f(r)-2k$. The curvature scalar satisfies $C^2=\mathcal W^2/(3r^4)$, so that
	\begin{equation}
		\partial_\theta C^2
		=\frac{2\mathcal W(r)}{3r^4}
		\left[r^2\partial_\theta f''(r)
		-2r\partial_\theta f'(r)
		+2\partial_\theta f(r)\right],
		\qquad \theta\in\{\qg,\lBI\}.
		\label{eq:weylWeightResponse}
	\end{equation}
	The Weyl response depends on $f$ and its radial derivatives. Its sign is determined by the full expression in Eq.~\eqref{eq:weylWeightResponse}, not by the signs of Eqs.~\eqref{eq:dqf} and \eqref{eq:dlf} alone. The inner Weyl potentials in Figure~\ref{fig:lambdaFixedPotential}(c,f) differ substantially across the scans. Their behavior near $r=0$ is explained by the powers found in Section~\ref{sec:weights}, not by the derivative formula alone. Taken together, the three scalar calculations show why charge and deformation can act differently on the weighted potentials.

\begin{figure}[!htbp]
		\centering
		\includegraphics[width=\linewidth]{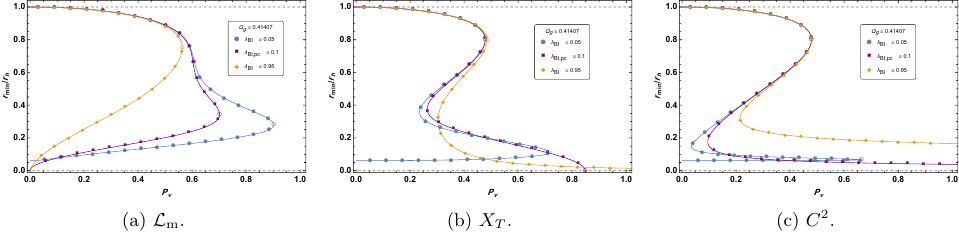}
		\caption{Planar momentum--turning-radius maps at fixed
			$\qg=0.41407$ for the three Born--Infeld deformation scans.
			The differences among inner candidate segments reflect their scalar-dependent potential structures.}
		\label{fig:lambdaFixedPenetration}
	\end{figure}

	Having examined how the Born--Infeld deformation modifies the effective potential, it is useful to investigate its effect on the turning radii of the extremal hypersurfaces. Figure~\ref{fig:lambdaFixedPenetration} shows the corresponding $r_{\min}$--$P_v$ curves for the effective potentials presented in Figure~\ref{fig:lambdaFixedPotential}(d)--(f). Varying the Born--Infeld coupling substantially changes the deeper turning radii, while some outer branches remain relatively close to each other.
	
	\section{Complexity as an interior probe}
	\label{sec:complexity}
	
	The turning-point maps tell us which extremal hypersurfaces can reach the AdS boundaries. We now calculate their complexities using Eq.~\eqref{eq:complexity}. At each time, the prescription \eqref{eq:envelope} selects the largest accessible on-shell value. We first study the growth of individual branches and then ask when two branches could exchange dominance.

	\subsection{Growth along accessible branches}
	
	Before the first change of dominant branch, the complexity grows along the ordinary outer turning-point family. Integrating the Hamilton--Jacobi relation \eqref{eq:HJ}, with the same $\Delta C(0)=0$ subtraction in each background, gives the exact identity
	\begin{equation}
		\Delta C^{\rm(out)}(x;\qg,\lBI)
		=\int_0^x P_{\rm out}(s;\qg,\lBI)\,\dd s,
		\label{eq:earlyIntegral}
	\end{equation}
	where $x=\tau/\LAdS$. Consequently, the parameter responses at fixed boundary time satisfy
	\begin{equation}
		\partial_\theta\Delta C^{\rm(out)}(x)
		=\int_0^x\partial_\theta P_{\rm out}(s)\,\dd s,
		\qquad \theta\in\{\qg,\lBI\}.
		\label{eq:earlySensitivity}
	\end{equation}
	Thus the early-time complexity responds to charge or deformation through the momentum $P_{\rm out}(x)$ at the same boundary time. The planar $P_v$--$r_{\min}$ scans offer a first indication: the outer folds usually move more when charge changes than when Born--Infeld deformation changes. At a smooth fold $r_f$, we have $U_X'(r_f)=0$ and $P_f^2=U_X(r_f)$. Differentiating gives
	\begin{equation}
		\frac{\dd P_f}{\dd\theta}
		=\frac{\partial_\theta U_X(r_f)}{2P_f},
		\qquad \theta\in\{\qg,\lBI\}.
		\label{eq:foldMomentumResponse}
	\end{equation}
	Eq.~\eqref{eq:foldMomentumResponse} gives the change of the critical momentum at a smooth fold. It does not determine how the momentum varies at an arbitrary early boundary time. That requires the full branch function $P_{\rm out}(x)$.
	
\begin{figure}[!htbp]
		\centering
		\includegraphics[width=\linewidth]{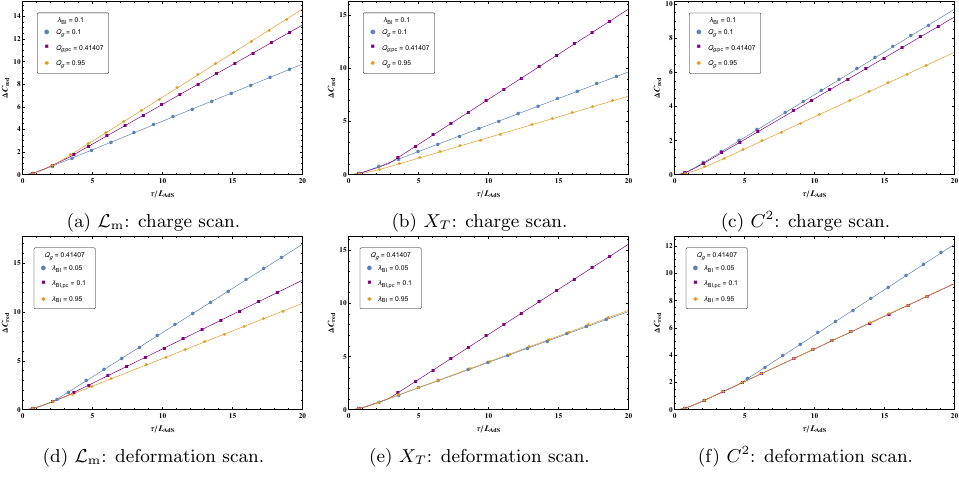}
		\caption{Planar complexity for the three scalar prescriptions.
			The upper row varies charge at fixed $\lBI=0.1$;
			the lower row varies $\lBI$ at fixed $\qg=0.41407$.
			Curves extending to $\tau/\LAdS\simeq20$ include
			numerical-model continuations. Sharp modeled changes
			are not, by themselves, verified equal-time saddle
			crossings.}
		\label{fig:complexityPlanar}
	\end{figure}

	Figure~\ref{fig:complexityPlanar} compares the planar complexity curves. During their initial growth, the charge scans generally separate more than the Born--Infeld deformation scans. The order of the curves depends on the scalar weight and on the parameter range. We compare the two responses more precisely by defining the ratio for equal infinitesimal fractional parameter changes:
	\begin{equation}
		R_P(x)=
		\frac{
			|\lBI\partial_{\lBI}P_{\rm out}(x)|
		}{
			|\qg\partial_{\qg}P_{\rm out}(x)|
		}.
		\label{eq:earlySensitivityRatio}
	\end{equation}
	
	Here $R_P(x)$ compares equal fractional changes of the two parameters at the same boundary time. If $R_P(x)<1$, the outer-branch momentum is less sensitive to Born--Infeld deformation than to charge.
	
	In our numerical analysis, we find that $R_P(x)<1$ in the vast majority of the sampled cases. The early-time outer-branch momentum is therefore predominantly more sensitive to fractional variations in the dressed charge than to equal fractional variations in the Born--Infeld deformation. Exceptions occur only for the matter Lagrangian prescription, whereas the inequality holds throughout the sampled charged configurations for the stress-trace and Weyl prescriptions. These numerical findings are consistent with the geometric response established in Eq.~\eqref{eq:geometricResponseBound}.

\begin{figure}[!htbp]
		\centering
		\includegraphics[width=\linewidth]{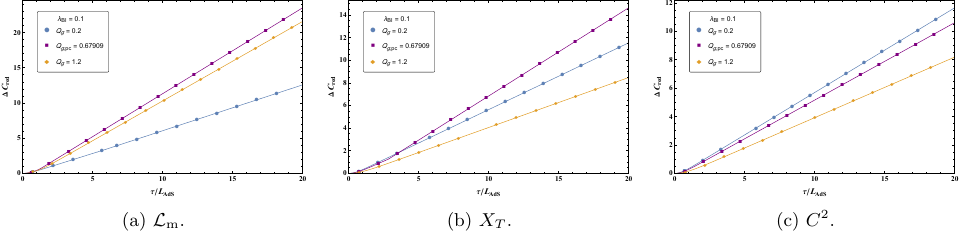}
		\caption{Spherical complexity for the fixed-$\lBI$ charge scans.
			The three scalar prescriptions exhibit different charge-dependent orderings because they weight the interior differently.}
		\label{fig:complexitySpherical}
	\end{figure}

	Figure~\ref{fig:complexitySpherical} shows the spherical results. The ordering of the curves depends on which scalar we use, consistent with the potential barriers in Figure~\ref{fig:UeffSpherical}. Each curve represents a different charge background; these curves are not competing branches of one black hole.
	
	When the turning radius of an accessible hypersurface approaches a nondegenerate maximum $r_f$ of the effective potential $U_X$, the boundary time diverges and the conserved momentum approaches the critical value $P_\infty=\sqrt{U_X(r_f)}$. The Hamilton--Jacobi relation then implies that the complexity grows linearly at late times:
	\begin{equation}
		\lim_{x\to\infty}\frac{\dd\Delta C}{\dd x}=P_\infty,
		\qquad
		\Delta C(x)=P_\infty x+O(1).
		\label{eq:criticalSlope}
	\end{equation}

\begin{figure}[!htbp]
		\centering
		\includegraphics[width=\linewidth]{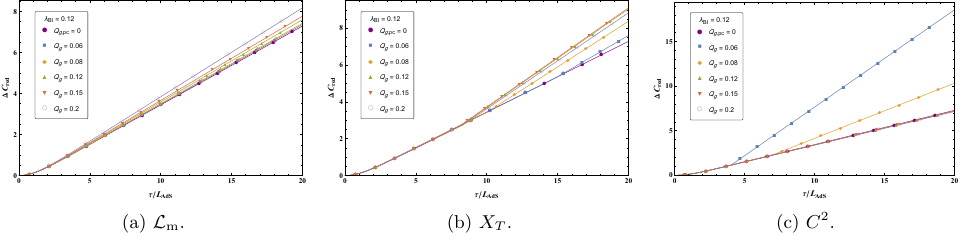}
		\caption{Hyperbolic complexity at fixed $\lBI=0.12$.
			The neutral case gives a common unweighted control.
			In the charged cases, candidate-branch competition depends
			on accessibility and equal-time complexity, not only on the
			count of local potential maxima. Extended-time results
			contain modeled continuations.}
		\label{fig:complexityHyperbolic}
	\end{figure}

	The neutral hyperbolic black hole provides a useful analytic check. When $k=-1$ and $\qg=0$, all three scalars vanish and the weights become unity. In our normalization,
	\begin{equation}
		U_X(r)=r^4(1-r^2),\qquad
		r_f=\sqrt{\frac23},\qquad
		P_\infty=\frac{2}{3\sqrt3}.
		\label{eq:hyperbolicbaseline}
	\end{equation}
	Charge changes these common results. The three weighted potential barriers then differ, as seen in Figures~\ref{fig:UeffHyperbolic} and \ref{fig:branchesHyperbolic}. Figure~\ref{fig:complexityHyperbolic} compares the resulting complexity curves with the neutral reference.
	
	\subsection{Competition between accessible branches}
	
	Competition between extremal hypersurfaces and the associated turning times have been studied in generalized volume-complexity \cite{WangJiangLiu2023,JiangWangLiu2024,WangLiWang2023}. Here we express the equality of two branch complexities in terms of their initial gap and integrated momentum difference. When multiple accessible hypersurfaces exist at the same boundary time, their on-shell complexities determine the dominant surface. To derive the condition for an exchange of dominance, consider two branches in the same black-hole geometry: an outer branch with complexity $C_{\rm out}(x)$ and a deeper branch with complexity $C_{\rm deep}(x)$. Suppose the outer branch has the larger complexity at some time $x_0$, with an initial complexity difference $D_0=C_{\rm out}(x_0)-C_{\rm deep}(x_0)>0$. Each branch obeys the Hamilton--Jacobi relation \eqref{eq:HJ}. Subtracting the two complexities gives
	\begin{equation}
		C_{\rm deep}(x)-C_{\rm out}(x)
		=-D_0+\int_{x_0}^{x}
		\big[P_{\rm deep}(s)-P_{\rm out}(s)\big]\dd s.
		\label{eq:branchDifference}
	\end{equation}
	The two branches can exchange dominance only when their on-shell complexities become equal. At a crossing time $x_*$, this requires
	\begin{equation}
		\int_{x_0}^{x_*}
		\big[P_{\rm deep}(s)-P_{\rm out}(s)\big]\dd s=D_0.
		\label{eq:crossingIntegral}
	\end{equation}
	The meaning is straightforward: the deeper branch must grow enough to overcome its initial complexity disadvantage $D_0$. If the momentum difference is nearly constant between $x_0$ and $x_*$, the crossing time is approximately
	\begin{equation*}
		x_*\simeq x_0+\frac{D_0}{P_{\rm deep}-P_{\rm out}}.
	\end{equation*}
	For the same initial gap, a larger positive momentum difference can make the crossing occur earlier. The critical momentum of one branch is not enough to determine the crossing time. Both branches must be accessible and their on-shell complexities must be compared at the same time. An exchange of dominance further requires the complexity difference to change sign across the crossing, and the two branches must attain the maximum in Eq.~\eqref{eq:envelope} on their respective sides of the crossing.
	
	\section{Conclusions}
	\label{sec:conclusion}
	We studied how three CEA weights probe four-dimensional GBI-AdS black holes with spherical, planar, and hyperbolic horizons. For a fixed black-hole geometry, we compared the on-shell electromagnetic Lagrangian, the negative stress-tensor trace, and the square of the Weyl tensor. These weights give different importance to different parts of the same interior.
	
	We first distinguished pole cancellation from regularity. Canceling the leading $1/r$ term in the blackening function does not remove the curvature singularity of a charged solution. It does, however, change the weighted effective potentials. At charged pole cancellation, the matter potential tends to zero, the stress-trace potential approaches a finite positive endpoint value, and the Weyl potential diverges. In the examined backgrounds, the stress-trace endpoint forms a one-sided maximum and gives a limiting turning-point momentum. Its contribution to complexity is subject to the same boundary-accessibility test as other turning points.
	
	The effective potentials and turning-radius maps reveal how the dressed charge, Born--Infeld deformation, and horizon topology influence the structure of the extremal hypersurface branches. At fixed horizon radius, we analytically established that, when $\lBI\qg^2/(8r_0^4)<1$, the blackening function is less sensitive to fractional variations in the Born--Infeld deformation than to equal fractional variations in the dressed charge over the corresponding radial interval. Our numerical analysis reveals a similar behavior for the early-time outer-branch momentum in the vast majority of the sampled cases, with exceptions occurring only for the matter Lagrangian prescription.
	
	When two accessible branches coexist at the same boundary time, their on-shell complexities determine which branch dominates. We derived an exact relation governing the competition between these branches in terms of their momentum difference: the branch that starts with smaller complexity must gain enough to close the initial gap, as expressed by Eq.~\eqref{eq:crossingIntegral}. The crossing time therefore depends on both the initial complexity difference and the subsequent momentum evolution of the two branches, rather than on the critical momentum of either branch alone. The extended-time figures include modeled continuations and should not be taken as direct evidence of a branch transition.
	
	Our results suggest several directions for future research. An important question is whether CEA observables can probe the dynamical formation of singularities and the instability of inner horizons in charged black holes. Another interesting direction is to investigate the role of electromagnetic duality in holographic complexity and to determine whether different scalar prescriptions can distinguish electromagnetic configurations that share the same spacetime geometry. It would also be interesting to explore the scaling behavior of these observables near extremality and its connection to the near-horizon geometry. More generally, one may ask whether different CEA observables can be used to reconstruct the interior geometry and matter content of black holes, and whether they provide a universal classification of black-hole interiors based on the causal nature and scaling properties of their singularities.
	
	\section*{acknowledgments}
	The author is grateful to Song He, Jorge G. Russo, Dmitri Sorokin, and Roberto Tateo for their interest in this work and the fruitful discussions that followed. H.B.-A. would like to express my sincere gratitude to Robert Myers and Shanming Ruan for the highly productive discussions during the "Quantum Information in Quantum Gravity 2026 (QIQG 2026)" conference in YMSC, Tsinghua University, Beijing. This work was conducted as part of the PostDoc Program on {\it Exploring TT-bar Deformations: Quantum Field Theory and Applications}, sponsored by Ningbo University. This work is based upon research funded by Iran National Science Foundation (INSF) under project No. 40509454

\end{document}